\documentclass[conference]{IEEEtran}
\IEEEoverridecommandlockouts
\usepackage{cite}
\usepackage{amsmath,amssymb,amsfonts}
\usepackage{dblfloatfix}    % To enable figures at the bottom of page
\usepackage{graphicx}
\usepackage{textcomp}
\usepackage{xcolor}
\usepackage{comment}
\usepackage{algorithm}
\usepackage{algpseudocode}
\usepackage{float}
\usepackage{bbm}
\def\BibTeX{{\rm B\kern-.05em{\sc i\kern-.025em b}\kern-.08em
		T\kern-.1667em\lower.7ex\hbox{E}\kern-.125emX}}
\usepackage{caption}	
\usepackage{tikz,pgf}
\usetikzlibrary{patterns}
\usetikzlibrary{calc}

\usepackage{pgfplots}

\newtheorem{theorem}{Theorem}

\newtheorem{definition}{Definition}

\begin{document}

\title{Virtual Cell Clustering with Optimal Resource Allocation to Maximize Cellular System Capacity
\thanks{This research was supported by Huawei, by AFOSR Grant FA9550-12-1-0215, and by ONR Grants N00014-15-1-2527 and N00014-18-1-2191.}}

\author{\IEEEauthorblockN{Michal Yemini}
\IEEEauthorblockA{Stanford University}
\and
\IEEEauthorblockN{Andrea J. Goldsmith}
\IEEEauthorblockA{Stanford University}
}

\maketitle

\begin{abstract}
This work presents a new network optimization framework for cellular networks using neighborhood-based optimization. Under this optimization framework resources are allocated within virtual cells encompassing several base-stations and the users within their coverage areas.  We form the virtual cells using hierarchical clustering with a minimax linkage criterion given a particular number of such cells. Once the virtual cells are formed, we consider an interference coordination  model in which  base-stations in a virtual cell jointly allocate the channels and power to users within the virtual cell. We propose two new schemes for solving this mixed integer NP-hard resource allocation problem. The first scheme transforms the problem into a continuous variables problem; the second scheme proposes a new channel allocation method and then alternately solves the channel allocation problem using this new method, and the power allocation problem. We evaluate the average system sum rate of these schemes for a variable number of virtual cells. These results quantify the sum-rate along a continuum of fully-centralized versus fully-distributed optimization for different clustering and resource allocation strategies. These results indicate that the penalty of fully-distributed optimization versus fully-centralized (cloud RAN) can be as high as 50\%. However, if designed properly,  a few base stations within a virtual cell using neighborhood-based optimization have almost the same performance as fully-centralized optimization.
\end{abstract}

%\begin{IEEEkeywords}
%component, formatting, style, styling, insert
%\end{IEEEkeywords}

\section{Introduction}
The  demand for increased capacity in cellular networks continues to grow, and is a major driver in the deployment of 5G systems. To increase cellular network capacity, the deployment of small cells has been proposed and is currently taking place  \cite{4623708,6768783,6171992,anpalagan_bennis_vannithamby_2015}. The proximity of  small cells to one another combined with their frequency reuse can cause severe interference to neighboring small cells and macrocells; this interference must be managed carefully to maximize the overall network capacity. Thus, powerful interference mitigation methods as well as optimal resource allocation schemes must be developed for 5G  networks. In this work we investigate a flexible resource allocation structure for cellular systems where, instead of each base-station serving all users within its own cell independently, several base-stations act cooperatively to create a “virtual cell” with joint resource allocation.
In order to design wireless networks that are composed of virtual cells we address in this work the following two design challenges: 1) Creating the virtual cells, i.e., cluster the base-stations and users into virtual cells. 2) Allocating the power and channels in each virtual cell. In this work we address the uplink resource allocation problem for joint channel and power allocation for the single user detection scenario.
This  resource allocation problem  is a non-convex NP hard problem even in  non-cooperative setups.

Base-station and user clustering as part of a resource allocation is discussed in the Cooperative Multi-Point (CoMP) literature, see for example  \cite{6530435,6707857,6181826,6555174,5594575,5502468,6655533,4533793,5285181,6786390,8260866}. The work \cite{8260866} presents an extensive literature survey of cell clustering for CoMP in wireless networks in which the clustering of base-stations and users is categorized as follows: 1) Static clustering which considers a cellular network whose cells are clustered statically, and does not adapt to network changes. Examples for static clustering algorithms are presented in \cite{6530435,6181826,6707857,6555174}. 2) Semi-dynamic clustering, in which static clusters are formed but the cluster affiliation of users is adapted according to changes in the network. Examples for such algorithms are presented in  \cite{5594575,5502468,6655533}. 3) Dynamic clustering in which the clustering of both base-stations and users adapts to changes in the network. Examples for dynamic clustering algorithms are presented in \cite{4533793,5285181,6786390}.
In addition, resource allocation in virtual cells for cooperative multi-point decoding is closely related to cloud radio access network design \cite{5594708,CIT-048,6924850,7487951,6601765} in which several base-stations act cooperatively. The coordination between the base-stations can be divided into the following categories: 1) Interference coordination in which only channel states are available at the coordinated base-stations. 2) Full cooperation in which base-stations share not only channel states but also full data signals they receive. 3) Rate limited coordination in which the base-stations are linked by limited-capacity backhaul. 4) Relay-assisted cooperation in which cooperation is carried by relays instead of  dedicated backhaul links. Our work  \cite{YeminiGoldsmith2} considers the full cooperation model in which  base-stations jointly decode their messages assuming infinite capacity backhaul links between base-stations in the same virtual cell. This manuscript considers a lower complexity and lower backhaul capacity scenario than \cite{YeminiGoldsmith2}, such that base-stations within a virtual cell share channel states via interference coordination  with single user detection rather than joint detection of all users in the virtual cell.

\paragraph*{Main Contributions}
This work considers cooperation across base-stations in a cellular network while preserving  desirable properties such as simple user association rules and low-complexity coordination within virtual cells  to suppress interference.
The virtual cell design and associated resource allocation is aimed at
1)	improving network performance while balancing  computational complexity,
2)	taking advantage of both local and global network information, and,
3)	ensuring that local changes in the network do not cause a ``butterfly effect" in which the whole virtual cell design and resource allocation must be recalculated.
We propose using hierarchical clustering to cluster base-stations since it enjoys the unique property that decreasing or increasing the number of clusters affects only the clusters that are being merged or separated, leaving all other clusters unchanged. We also propose two new resource allocation schemes for virtual cells with multiple BSs, which   is a mixed-integer NP-hard problem. The first scheme converts the problem into a continuous variable problem, whereas  the second scheme proposes a new channel allocation scheme and then alternates between allocating  channels using this new channel allocation scheme and allocating  power. Our numerical results show that our new resource allocation schemes for virtual cells outperform previously proposed resource allocation solutions in a fully centralized setup.

\section{Problem Formulation}\label{sec:problem_formualtion}
We consider  a communication network that comprises a set of base-stations (BSs) $\mathcal{B}$, a set of users $\mathcal{U}$ and a set of frequency bands $\mathcal{K}$. The users communicate with the BSs and interfere with the transmissions of one another. Each user $u\in\mathcal{U}$ has a power constraint of $\overline{P}_u$ dBm.
The BSs and users are clustered into virtual cells which must fulfill the following characteristics.

\subsection{Virtual Cells}\label{sec:virtual_cell_requirements}

\begin{definition}[Virtual BS]
	Let $b_1,..,b_n$ be $n$ BSs in a communication network, we call the set $\{b_1,..,b_n\}$ a virtual BS.
\end{definition}
\begin{definition}[Proper clustering]	
	Let $\mathcal{B}$ be a set of BSs,  $\mathcal{U}$ be a set of users. Denote   $\mathcal{V}=\{1,\ldots,V\}$.
	For every $v$, define the sets $\mathcal{B}_v\subset \mathcal{B}$ and $\mathcal{U}_v\subset \mathcal{U}$ .
	We say that the set $\mathcal{V}$ is a proper clustering of the sets  $\mathcal{B}$ and  $\mathcal{U}$  if $\mathcal{B}_v$ and $\mathcal{U}_v$ are partitions of the sets $\mathcal{B}$ and $\mathcal{U}$, respectively. That is,
	$\bigcup_{v\in\mathcal{V}}\mathcal{B}_v = \mathcal{B}$, $\bigcup_{v\in\mathcal{U}}\mathcal{U}_v = \mathcal{U}$. Additionally,
	$\mathcal{B}_{v_1}\cap\mathcal{B}_{v_2}=\emptyset$ and $\mathcal{U}_{v_1}\cap\mathcal{U}_{v_2}=\emptyset$ for all $v_1,v_2\in\mathcal{V}$ such that $v_1\neq v_2$.
\end{definition}

\begin{definition}[Virtual cell]
	Let $\mathcal{B}$ be a set of BSs, $\mathcal{U}$ be a set of users, and  $\mathcal{V}$ be a proper clustering of $\mathcal{B}$ and $\mathcal{U}$. For every $v\in\mathcal{V}$  the virtual cell  $\mathcal{C}_v$ is composed of the virtual BS $\mathcal{B}_v$ and the set of users $\mathcal{U}_v$.	
\end{definition}

This condition ensures that every BS and every user belongs to exactly one virtual cell. This implies that all the transmission power of a user is dedicated to communicating with base-stations in the same virtual cell, thus power allocation can be optimized in a virtual cell.

Let $\mathcal{V}$ be a proper clustering of the set of BSs $\mathcal{B}$ and the set of users $\mathcal{U}$, and let   $\{\mathcal{C}_v\}_{v\in\mathcal{V}}$ be the set of virtual cells that $\mathcal{V}$ creates.
In each virtual $\mathcal{C}_v$ we assume that the BSs that compose the virtual BS $\mathcal{B}_v$ jointly allocate their resources.

\subsection{The Uplink Resource Allocation Problem  }\label{subsection:uplink_interference_coordination_problem}

In each virtual cell we consider the uplink resource allocation problem in which  the BSs in the virtual cell jointly optimize the channel allocation and the transmission power  of the users within it. We  consider  single user detection in which every BS $b$ decodes each codeword associated with a given user separately (i.e. there is no multiuser detection).

Recall that $\mathcal{K}$ is the set of frequency bands.
Denote by $h_{u,b,k}$  the channel coefficient of the channel from user $u\in\mathcal{U}$ to BS $b$ over frequency band $k$, and let $P_{u,k}$ be the transmit power of user $u$ over frequency band $k$. Further, let $\sigma^2_{b,k}$ denote the noise power at BS $b$ over frequency band $k$, and let $W_k$ denote the bandwidth of band $k$.
The uplink resource allocation problem in a virtual cell $\mathcal{C}_v$, ignoring  interference from other virtual cells, is given by:
\begin{flalign}\label{eq:no_decoding_cooperation_single_discrete}
\max & \sum_{b\in\mathcal{B}_v}\sum_{u\in\mathcal{U}_v}\sum_{k\in\mathcal{K}}
\gamma_{u,b,k}W_k\log_2\left(1+\frac{|h_{u,b,k}|^2P_{u,k}}{\sigma^2_{b,k}+J_{u,b,k}}\right)\nonumber\\
\text{s.t.: } & \quad 0\leq P_{u,k},\hspace{0.8cm} \sum_{k\in\mathcal{K}}P_{u,k} \leq \overline{P}_u,\quad \forall\: u\in \mathcal{U}_v,k\in\mathcal{K}\nonumber\\
%&  \sum_{u\in\mathcal{U}_v} g_{u,\tilde{b}}P_{u}\leq \ J_{v,\tilde{b}},\quad \forall \: \tilde{b}\notin v,\nonumber\\
&\hspace{-0.15cm} \sum_{\substack{\tilde{u}\in\mathcal{U}_v, \tilde{u}\neq u}} |h_{\tilde{u},b,k}|^2P_{\tilde{u},k}=  J_{u,b,k},\: \forall  u\in\mathcal{U}_v,b\in \mathcal{B}_v,k\in\mathcal{K} \nonumber\\
& \sum_{b\in\mathcal{B}_v}\gamma_{u,b,k}\leq 1 ,\quad \forall\:u\in \mathcal{U}_v,k\in\mathcal{K}\nonumber\\
& \gamma_{u,b,k}\in\{0,1\},\quad \forall\:u\in \mathcal{U}_v, b\in\mathcal{B}_v.
\end{flalign}
This is a mixed-integer programming problem that is NP-hard.

We note that our approach mitigates interference in the network by merging cells to create virtual cells. While currently we do not mitigate interference between virtual cells,  as their number is decreased, they become larger and the interference inside the virtual cells becomes the dominant interference.
This interference is mitigated in  (\ref{eq:no_decoding_cooperation_single_discrete})  to improve network performance. Numerical results presented in Section \ref{sec:numerical_results} indeed confirm the improvement in network average sum rate that our method provides.

\section{Forming the Virtual Cells}\label{sec:virtual_cell_create}
This section presents the clustering approaches that we develop to create the virtual cells within which the resource allocation schemes we present in Sections \ref{sec:joint_power_allocation}-\ref{sec:alternating_optimization}  operate.
\subsection{Base-Station Clustering}
We propose using  the hierarchical clustering algorithm with minimax linkage \cite{BienTibshirani2011} to cluster BSs since it enjoys the unique property that decreasing or increasing the number of clusters  only affects the clusters that are being merged or separated. Thus, the number of clusters can adapt efficiently to the current state of the network without requiring a full clustering update. By contrast, in other clustering methods such as K-means or spectral clustering, even a small variation in the number of clusters requires the recalculation of all the clusters in the network. This is undesirable in wireless networks since it inflicts a large setup time overhead for each reclustering that is caused by the need for information acquisition and other message passing.

The agglomerative hierarchical clustering algorithm using the minimax linkage criterion is presented in Algorithm \ref{algo:hierarchical_clustering}; it gets  a set of points $S$ and produces the clusterings $B_1,\ldots,B_n$, where $B_m$ is the clustering of size $m$. The algorithm defines the center of a cluster to be the member of the cluster with the minimal maximal distance to all other members in the cluster; this minimal maximal distance is the cluster radius. Then, in every step the minimax linkage criterion merges the two clusters that will jointly have the smallest radius out of all merging possibilities.

Let $d:\mathbb{R}^2\times\mathbb{R}^2\rightarrow\mathbb{R}$ be the Euclidean distance function, and let $S$ be a set of points in $\mathbb{R}^2$. We then define the following:
\begin{definition}[Radius of a set around point]
	The radius of $S$ around $s_i \in S$  is defined as $r(s_i,S)=\max_{s_j\in S}\:d(s_i,s_j)$.
\end{definition}
\begin{definition}[Minimax radius]
	The minimax radius of $S$ is defined as $r(S) = \min_{s_i\in S}\: r(s_i,S)$.
\end{definition}
\begin{definition}[Minimax linkage]
	The minimax linkage between two sets of points $S_1$ and $S_2$ in $\mathbb{R}^2$ is defined as $d(S_1,S_2) = r(S_1\cup S_2)$.
\end{definition}

Let $S=\{s_1,\ldots,s_n\}$ be the set of locations of the BSs in $\mathcal{B}$. We use Algorithm \ref{algo:hierarchical_clustering} below with the input $S$ to create the  virtual BSs for each number of clusters $m$; this produces the dendrogram which shows what clusters are merged as the number of clusters is decreased.

We chose to use the minimax linkage criterion since interference tends to increase on average as the distance between interferers is decreased. Thus, at each stage the minimax linkage criterion  merges the two clusters of base-stations that maximize the smallest anticipated interference at the center of the new cluster that is caused by base-stations in the cluster.  In addition, the minimax linkage criterion benefits from fulfilling several desirable properties in cluster analysis as discussed in  \cite{BienTibshirani2011}.

\vspace{-0.1cm}
\begin{algorithm}
	\caption{}\label{algo:hierarchical_clustering}
	\begin{algorithmic}[1]		
		\State Input: A set of point $S=\{s_1,\ldots,s_n\}$;
		\State Set $B_n = \left\{\{s_1\},\dots,\{s_n\}\right\}$;
		\State Set $d(\{s_i\},\{s_j\})=d(s_i,s_j),\:\forall s_i,s_j\in S$;	
		\For {$m = n-1,\ldots,1$}
		\State Find $(S_1,S_2) = \arg\min_{\stackrel{G,H\in B_{m+1}:}{G\neq H}} d(G,H)$;
		\State  Update $B_{m} = B_{m+1} \bigcup \{S_1\cup S_2\} \setminus \{S_1,S_2\}$;
		\State Calculate $d(S_1\cup S_2,G)$ for all $G\in B_m$;
		\EndFor
		%\EndProcedure
	\end{algorithmic}
	
\end{algorithm}
\vspace{-0.1cm}

\subsection{Users' Affiliation with Clusters}\label{sec_user_affil}
To create the virtual cells, we consider two affiliation rules: 1) Closest BS rule in which each user is affiliated with its closest BS. 2) Best channel rule in which  each user is affiliated with the BS to which it has the best channel  (absolute value of the channel coefficient). Then each user is associated with the virtual BS that its affiliated BS is part of.
This way every virtual BS and it associated users compose a virtual cell.

It is easy to verify that this formation of the virtual cells fulfills the requirement presented in Section \ref{sec:virtual_cell_requirements}.

\section{Joint Channel  and Power  Allocation}\label{sec:joint_power_allocation}
This section introduces the first  resource allocation scheme we propose. It is found by converting the problem (\ref{eq:no_decoding_cooperation_single_discrete}) to an equivalent continuous variable problem and then solving it via an approximation.
\subsection{An Equivalent Continuous Variable  Problem}
We can represent the problem (\ref{eq:no_decoding_cooperation_single_discrete}) by an equivalent problem with continuous variables. Suppose that instead of sending a message to at most one single BS at each frequency band, a user sends messages to all BSs. The signal of user $u\in\mathcal{U}_v$ over frequency band $k$ is then given by $x_{u,k}=\sum_{b\in\mathcal{B}_v}x_{u,b,k}$ where $x_{u,b,k}$ is the part of the signal of user $u$ that is transmitted over frequency band $k$ and is  intended to be decoded by BS $b$. Let $P_{u,b,k}$ be the power allocation of the part of the signal of user $u$ that is transmitted over frequency band $k$ and is  intended to be decoded by BS $b$.; i.e. $P_{u,b,k}=E\left( x_{u,b,k}^2\right)$, where $E\left( x_{u,b,k}^2\right)$ denotes the expected value of $x_{u,b,k}^2$.
We next prove that (\ref{eq:no_decoding_cooperation_single_discrete}) can in fact be written in the following equivalent form:
\begin{flalign}\label{eq:no_decoding_cooperation_single_continuous}
\max & \sum_{b\in\mathcal{B}_v}\sum_{u\in\mathcal{U}_v}\sum_{k\in\mathcal{K}}
W_k\log_2\left(1+\frac{|h_{u,b,k}|^2P_{u,b,k}}{\sigma^2_{b,k}+J_{u,b,k}}\right)\nonumber\\
\text{s.t.: }
&\quad 0\leq P_{u,b,k},\hspace{1cm}  \sum_{b\in\mathcal{B}_v}\sum_{k\in\mathcal{K}} P_{u,b,k}\leq \overline{P}_u,\nonumber\\
& \hspace{-1cm}\sum_{\substack{(\tilde{u},\tilde{b})\in\mathcal{U}_v\times \mathcal{B}_v,\\(\tilde{u},\tilde{b})\neq (u,b)}}\hspace{-0.5cm} |h_{\tilde{u},b,k}|^2P_{\tilde{u},\tilde{b},k}=  J_{u,b,k},\: \forall\:  u\in\mathcal{U}_v,b\in \mathcal{B}_v,k\in\mathcal{K}.
\end{flalign}
\vspace{-0.1cm}

\begin{theorem}\label{theorem:equivalence:discrete_continuous}
	The mixed integer programming problem (\ref{eq:no_decoding_cooperation_single_discrete}) and the continuous variables problem (\ref{eq:no_decoding_cooperation_single_continuous})  are equivalent. 	
\end{theorem}

\begin{IEEEproof}
	The equivalence of (\ref{eq:no_decoding_cooperation_single_discrete}) and (\ref{eq:no_decoding_cooperation_single_continuous}) is argued as follows.
	
	First, the solution of (\ref{eq:no_decoding_cooperation_single_discrete}) can be achieved by the solution of (\ref{eq:no_decoding_cooperation_single_continuous}) by setting $x_{u,b,k}=0$ whenever $\gamma_{u,b,k}=0$, and $E \left(x_{u,b,k}^2\right) = P_{u,k}$ whenever $\gamma_{u,b,k}=1$; thus the maximal sum rate that is found by solving (\ref{eq:no_decoding_cooperation_single_continuous}) upper bounds the maximal sum rate  that is found by solving (\ref{eq:no_decoding_cooperation_single_discrete}).
	On the other hand, suppose that the optimal transmission power of user $u$ using frequency band $k$, given the transmission power of all other users, is $P_{u,k}$, that is $P_{u,k} = \sum_{b\in\mathcal{B}_v}P_{u,b,k}$.
	It follows by the duality between the multiple-access channel and the broadcast channel that is proved in  \cite{1237143} that the optimal power allocation $(P_{u,b,k})_{b\in\mathcal{B}_v}$ for user $u$ in frequency band $k$, given the power allocation of all other users, is to allocate all its transmission power $P_{u,k}$ over frequency band $k$ to the transmission to the BS with the highest SINR.
	It follows that the maximal sum rate of (\ref{eq:no_decoding_cooperation_single_continuous}) cannot be larger than that of (\ref{eq:no_decoding_cooperation_single_discrete}). Thus, the two problems (\ref{eq:no_decoding_cooperation_single_discrete}) and (\ref{eq:no_decoding_cooperation_single_continuous}) are equivalent.
\end{IEEEproof}
\subsection{Solving an Approximation of the Continuous Variable Resource Allocation Problem Optimally}\label{sec:continuous_HSINR_gradient}
In the following, we approximately solve the problem (\ref{eq:no_decoding_cooperation_single_continuous}). We note that since we solve a convex approximation of the problem, we may achieve a suboptimal solution.
Denote:
%\vspace{-0.02cm}
\begin{flalign}\label{eq:SINR_def}
\text{SINR}_{u,b,k}(\boldsymbol P) =\frac{|h_{u,b,k}|^2 P_{u,b,k}}{\sigma^2_b+\sum_{\substack{(\tilde{u},\tilde{b})\in\mathcal{U}_v\times \mathcal{B}_v,\\(\tilde{u},\tilde{b})\neq (u,b)}} |h_{\tilde{u},b,k}|^2P_{\tilde{u},\tilde{b},k}},
\end{flalign}
\vspace{-0.05cm}
where $\boldsymbol P = (P_{u,b,k})_{(u,b,k)\in\mathcal{U}_{v}\times\mathcal{B}_{v}\times\mathcal{K}}$ is the matrix of the transmission power.
Using the high SINR approximation \cite{5165179}
\begin{flalign*}
&\log(1+z)\geq \alpha(z_0)\log z+\beta(z_0),\nonumber\\
&\alpha(z_0) = \frac{z_0}{1+z_0}, \quad\beta(z_0) =\log(1+z_0)-\frac{z_0}{1+z_0}\log{z_0},
\end{flalign*}
 transforming the variables of the problem using $P_{u,b,k}=\exp(g_{u,b,k})$, and noticing that the terms $\beta_{u,b,k}^{(m)}$ do not affect the optimal power allocation, yield the following approximated convex iterative problem:
\begin{flalign}\label{sol_continuous_power_approx}
&\boldsymbol g^{(m)} = \arg\max \sum_{b\in\mathcal{B}_v}\sum_{u\in\mathcal{U}_v}\sum_{k\in\mathcal{K}}
W_k\alpha_{u,b,k}^{(m)}\nonumber\\
&\hspace{0.5cm}\cdot\log_2\left(\frac{|h_{u,b,k}|^2\exp(g_{u,b,k})}{\sigma^2_{b,k}+
	\sum_{\substack{(\tilde{u},\tilde{b})\in\mathcal{U}_v\times \mathcal{B}_v,\\(\tilde{u},\tilde{b})\neq (u,b)}} |h_{\tilde{u},b,k}|^2\exp(g_{\tilde{u},\tilde{b},k})}\right)\nonumber\\
&\text{s.t.: }    \sum_{b\in\mathcal{B}_v}\sum_{k\in\mathcal{K}} \exp(g_{u,b,k})\leq \overline{P}_u,\quad \forall\: u\in \mathcal{U}_v,
\end{flalign}
where $\alpha_{u,b,k}^{(m)}=\alpha(\text{SINR}_{u,b,k}(\boldsymbol P^{(m-1)}))$ and $\alpha_{u,b,k}^{(0)}=1$,  and $\boldsymbol g = (g_{u,b,k})_{(u,b,k)\in\mathcal{U}_{v}\times\mathcal{B}_{v}\times\mathcal{K}}$.

Since the problem (\ref{sol_continuous_power_approx}) is convex, its solution can be found by solving its dual problem using the gradient method following the analysis presented in \cite{5165179}. Furthermore, the convexity of (\ref{sol_continuous_power_approx}) implies that its Karush–Kuhn–Tucker (KKT) conditions are sufficient,  and a point that fulfills them is both primal and dual optimal.
Similar to the analysis presented in \cite{5165179}, if the iterative update rule (\ref{update_rule_continuous})  converges then it must converge to a KKT point, which in turn is globally optimal. While there is no known proof that guarantees convergence, in practice the following fixed point iteration converges:
\begin{flalign}\label{update_rule_continuous}
&P_{u,b,k}^{(m,s+1)}=\\
&\frac{W_k\alpha_{u,b,k}^{(m)}}{\lambda_u^{(s+1)}\ln 2\hspace{-0.08cm}+\hspace{-0.08cm}W_k\sum_{\substack{(\tilde{u},\tilde{b})\in\mathcal{U}_v\times \mathcal{B}_v,\\(\tilde{u},\tilde{b})\neq (u,b)}}\hspace{-0.1cm}\alpha_{\tilde{u},\tilde{b},k}^{(m)}\frac{\text{SINR}_{\tilde{u},\tilde{b},k}(\boldsymbol P^{(m,s)})|h_{u,\tilde{b},k}|^2}{P_{\tilde{u},\tilde{b},k}^{(m,s)}|h_{\tilde{u},\tilde{b},k}|^2 }}\nonumber
\end{flalign}
%\vspace{-0.4cm}
where $\lambda_u^{(s+1)}=0$ if
%\vspace{-0.2cm}
\begin{flalign*}\hspace{-0.07cm}
\sum_{b\in\mathcal{B}_v}\sum_{k\in\mathcal{K}}\hspace{-0.05cm}
\frac{\alpha_{u,b,k}^{(m)}}{\sum_{\substack{(\tilde{u},\tilde{b})\in\mathcal{U}_v\times \mathcal{B}_v,\\(\tilde{u},\tilde{b})\neq (u,b)}}\alpha_{\tilde{u},\tilde{b},k}^{(m)}\frac{\text{SINR}_{\tilde{u},\tilde{b},k}(\boldsymbol P^{(m,s)})}{P_{\tilde{u},\tilde{b},k}^{(m,s)}|h_{\tilde{u},\tilde{b},k}|^2, }|h_{u,\tilde{b},k}|^2}\leq \overline{P}_u,
\end{flalign*}
otherwise, it is chosen  such that $\sum_{b\in\mathcal{B}_v,k\in\mathcal{K}}P_{u,b,k}^{(m,s+1)}=\overline{P}_u$.

\section{Solving the Resource Allocation Problem  via Alternating Algorithms}\label{sec:alternating_optimization}

A more traditional approach for solving the problem  (\ref{eq:no_decoding_cooperation_single_discrete}) separates it into two subproblems: a channel allocation problem that sets the value of $\gamma_{u,b,k}$, and  a power allocation problem that optimizes the transmission power.  Then we iteratively solve these two problems until a stopping criterion is fulfilled.
A resource allocation scheme of this type is depicted in Algorithm \ref{algo:Altenating_general}.
\newline

\begin{algorithm}
	\caption{}\label{algo:Altenating_general}
	\begin{algorithmic}[1]
		%\Procedure{MyProcedure}{}
		\State Input: $\delta>0, N_{\max}\in\mathbb{N}$.
		\State Set $n=0$,  $\delta_0 = 2\delta$, $R_0 = 0$;
		\State Set $P^{(0)}_{u,k}=\overline{P}_u/|\mathcal{K}|$ for all $u\in\mathcal{U}_v$
		and $k\in\mathcal{K}$;
		\While{  $\delta_n>\delta$ and $n<N_{\max}$}
		\State $n=n+1$;
		\State \textbf{Channel allocation:} Given the power allocation $P^{(n-1)}_{u,k}$,  set $\gamma^{(n)}_{u,b,k}$ to be either zero or one for every $u\in\mathcal{U}_v$, $b\in\mathcal{B}_v$ and $k\in\mathcal{K}$.		
		\State \textbf{Power allocation:}  Given  $(\gamma_{u,b,k}^{(n)})_{(u,b,k)\in\mathcal{U}_v\times\mathcal{B}_v\times\mathcal{K}}$, calculate $(P_{u,b,k}^{(n)})_{(u,b,k)\in\mathcal{U}_v\times\mathcal{B}_v\times\mathcal{K}}$
		by solving the iterative problem (\ref{sol_continuous_power_approx}) starting with the
		initial values  $\alpha_{u,b,k}^{(0)}=\gamma_{u,b,k}^{(n)}$ for all $(u,b,k)\in\mathcal{U}_v\times\mathcal{B}_v\times\mathcal{K}$.
		\State Calculate: $P^{(n)}_{u,k}=\sum_{b\in\mathcal{B}_v}P_{u,b,k}^{(n)}$;
		\State Calculate the sum rate \[R_n\hspace{-0.1cm} =\hspace{-0.1cm}\sum_{b\in\mathcal{B}_v}\hspace{-0.05cm}\sum_{u\in\mathcal{U}_v}\hspace{-0.05cm}\sum_{k\in\mathcal{K}}
		\gamma^{(n)}_{u,b,k}W_k\log_2\left(1+\frac{|h_{u,b,k}|^2P^{(n)}_{u,b,k}}{\sigma^2_{b,k}+J^{(n)}_{u,b,k}}\right); \]
		\State Calculate $\delta_n = R_n-R_{n-1}$;
		\EndWhile
		%\EndProcedure
	\end{algorithmic}
\end{algorithm}

Next we present three channel allocation schemes for subproblem 1 that we consider in this work.

\subsection{User-Centric (UC) Channel Allocation}\label{sec:alternating_power_allocation_user}
In addition to the previously-proposed channel allocation schemes that we discuss in Section \ref{sec:alternating_power_allocation_BS}, we propose in Algorithm \ref{algo:Altenating_single_set_power_UCB}  a user-centric (UC) approach in which at each frequency band every user chooses its receiving BS to be the one  with the maximal SINR for this user given an initial power allocation.
The motivation behind this approach is allowing the power allocation stage more flexibility to choose the users who transmit to a given BS. More specifically,
in previously proposed channel allocation schemes discussed in Section \ref{sec:alternating_power_allocation_BS}, at most one user is allocated to a BS at each frequency band. However, in the UC approach at each frequency band each BS has a list of users that chose it as their receiving BS, then the power allocation stage chooses the identity of the user in that list who actually transmits to the BS by allocating to that user a positive transmission power.
Interestingly, numerical results show that as the number of virtual cells decreases and their size increases, both the UC channel allocation and the equivalent continuous problem approach outperform both of the previously-proposed channel allocation  methods that we next discuss.

\begin{algorithm}%[b]
	\caption{}\label{algo:Altenating_single_set_power_UCB}
	%\vspace{-0.1cm}
	\begin{algorithmic}[1]
		%\Procedure{MyProcedure}{}
		\State Input: Power allocation $\boldsymbol P = (P_{u,k})_{u\in\mathcal{U}_v,k\in\mathcal{K}}$;
		\State For every $u\in\mathcal{U}_v$, $b\in\mathcal{B}_v$ and $k\in\mathcal{K}$ calculate \[\text{SINR}_{u,b,k}(\boldsymbol P) = \frac{|h_{u,b,k}|^2P_{u,k}}{\sigma^2_{b,k}+\sum_{\tilde{u}\in\mathcal{U}_v ,\tilde{u}\neq u} |h_{\tilde{u},b,k}|^2P_{\tilde{u},k}};\]
		\State For every $u\in\mathcal{U}_v$ and $k\in\mathcal{K}$, calculate: \hspace{4cm} $b_{u,k} = \arg\max_{b\in\mathcal{B}_v} \text{SINR}_{u,b,k}(\boldsymbol P)$;
		\State For every $(u,b,k)\in\mathcal{U}_v\times\mathcal{B}_v\times\mathcal{K}$ set $\gamma_{u,b,k}=\mathbbm{1}_{\{b = b_{u,k}\}}$;
		%\EndProcedure
	\end{algorithmic}
\end{algorithm}

\subsection{Previously-Proposed Channel Allocation Schemes
}\label{sec:alternating_power_allocation_BS}
We also consider the following two existing channel allocation methods.  Due to space limitation we briefly discuss these existing schemes and refer the reader to the relevant works where they have been analyzed for more details.

\subsubsection{BS-centric (BSC) Channel Allocation}
In the BS-centric  (BSC) approach at each frequency band every BS chooses its transmitting user to be the one with the maximal SINR; this scheme is used in several  works such as \cite{6678362} and \cite{6815733}. We remark that we do not restrict BSs to choose from a designated subset of users, that is, they can choose to communicate with any user in the  virtual cell. We remark that even though theoretically in this approach several BSs can choose the same user, it can be proved following the argument presented in the proof of Theorem \ref{theorem:equivalence:discrete_continuous} that an optimal power allocation scheme  will allocate power only to the transmission of no more than one BS. In practice, this behavior is  observed using the high SINR approximation.

\subsubsection{Maximum Sum Rate Matching  (MSRM) Channel Allocation} \label{sec:channel_assignment}

The second channel allocation approach is presented in \cite{7873307}. It allocates the channels in a virtual cell optimally for a given power allocation by solving the maximum sum rate matching problem.  We note that, as stated in \cite{7873307}, given a power allocation $\boldsymbol P$, the maximum sum rate matching channel allocation approach finds the optimal channel allocation that maximizes the sum rate for that power allocation. However, since the power allocation may not be optimal, the overall solution is not necessarily optimal.

\section{Numerical results}\label{sec:numerical_results}
This section presents Monte Carlo simulation results for the system model presented in this paper.  There were $8$ frequency bands each of bandwidth  20 KHz.
The noise power received by each BS was $-174$ dBm/Hz, and the maximal power constraint for each user was $23$ dBm. In each frequency band we considered Rayleigh fading, Log-Normal shadowing with a standard deviation of $8$ dB and a path loss model of $PL(d)= 35\log_{10}(d)+34$ where $d$ denotes the distance between the transmitter and the receiver in meters (see \cite{4138008}).
The network was comprised of $10$ BSs and $80$ users which were uniformly located in a square of side $1000$ meters. The results were averaged over $1000$ system realizations.

\begin{figure}
	\vspace{-0.15cm}
	\centering
	\includegraphics[scale=0.6]{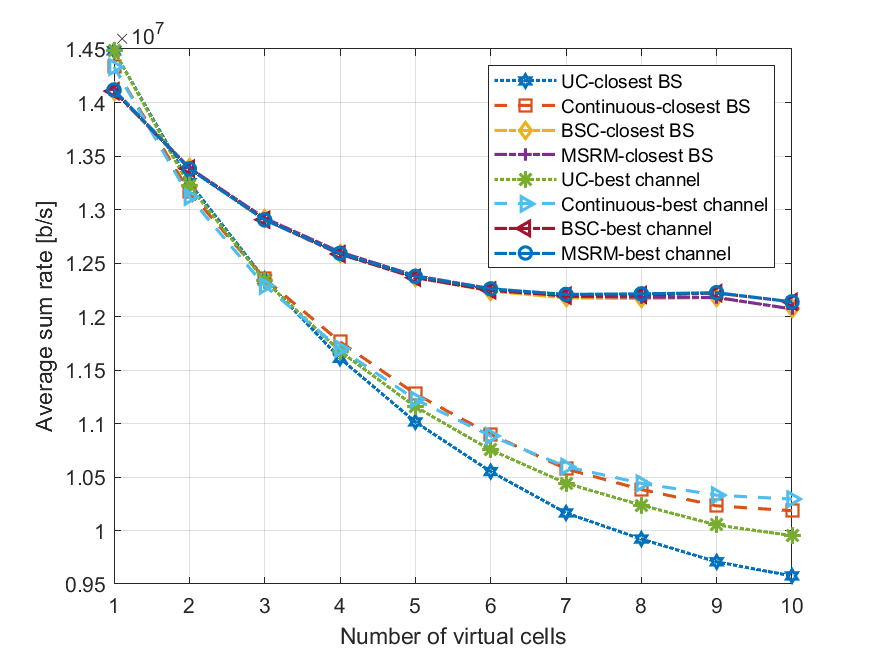}
	%,clip,trim=4cm 8.5cm 4.5cm 9cm
	\caption{Comparison of the average system sum rate as a function of the number of virtual cells that are created by the hierarchical clustering algorithm. The legend is written in the form X-Y where X and Y indicate the resource allocation scheme and  the user affiliation rule, respectively.}
	\label{hierarchical_sum_rate}
	\vspace{-0.5cm}
\end{figure}

Fig.~\ref{hierarchical_sum_rate} evaluates the average system sum rate as a function of the number of virtual cells for each of the resource allocation schemes we present. Fig.~\ref{hierarchical_sum_rate} leads to several interesting insights and conclusions. First, it confirms the expectation that as the number of virtual cells decreases, the average sum rate increases. Second, it shows that the best channel affiliation rule outperforms the closest BS one when the number of virtual cells is large. Third,
it compares the performance of the resource allocation schemes presented in this work. Fig.~\ref{hierarchical_sum_rate} shows that it is best to use the BSC or MSRM channel allocation methods, which yielded similar performance,  for allocating channels and power in virtual cells except when there is a single virtual cell (fully centralized optimization). In this case the two new resource allocation techniques that we propose outperform these other methods.
This can be explained by the fact that our new schemes  provide more freedom in the power allocation stage to choose which users have a positive transmission power compared with existing methods. However, since the power allocation problem is solved approximately, its solution may not be optimal. When the size of virtual cells is small, the channel allocation choice of the existing methods is good whereas the new methods suffer loss in performance due to the suboptimality of the power allocation stage. However, as the size  of the virtual cells grows (as their number is decreased), the ability of the new methods to consider in the power allocation stage more channel allocation combinations improves the resource allocation performance even though solution is only approximate.

\begin{figure}
	\vspace{-0.15cm}
	\centering
	\includegraphics[scale=0.62]{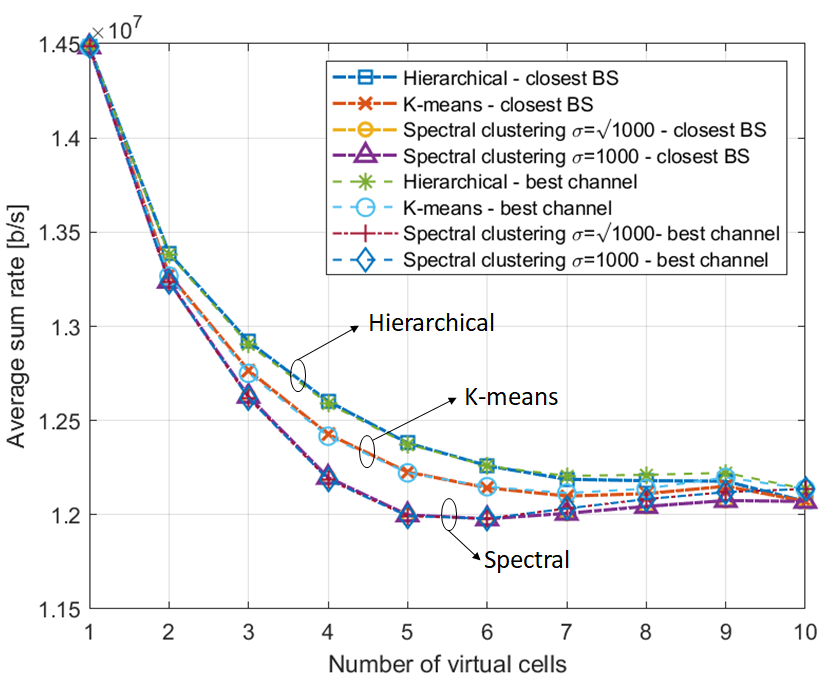}
	%,clip,trim=4cm 8.5cm 4.5cm 9cm
	\caption{Comparison of the average sum rate as a function of the number of virtual cells. The legend is written in the form X-Y where X and Y indicate the BS clustering algorithm and the user affiliation rule, respectively.}
	\label{Compare_clustering_algorithms}
	\vspace{-0.5cm}
\end{figure}

We also compared the average system sum rate that several BS clustering algorithms yielded, namely, hierarchical clustering with minimax linkage with that of the K-means clustering algorithm and that of the spectral clustering  algorithm \cite{Ng:2001:SCA:2980539.2980649} for the choices $\sigma=\sqrt{1000}$ and $\sigma=1000$.  Fig.~\ref{Compare_clustering_algorithms} depicts the   maximal average system sum rate achieved by each of the clustering algorithms where the maximization is taken over the resource allocation schemes considered in this work. Fig.~\ref{Compare_clustering_algorithms} shows that the hierarchical algorithm outperforms both the K-means and the spectral clustering algorithms. It also shows that a proper choice of  the clustering algorithm is crucial for improving network performance; this is evident in the plot of the spectral clustering algorithm in which the network performance monotonically decreases as the number of virtual cells is decreased from 10   to 5.
\section{Conclusion}
This work addressed the role of resource allocation and user affiliation within  virtual cells to maximize the sum rate of wireless networks. It proposed  solutions for two design aspects; namely, forming
the virtual cells and allocating the communication resources
in each virtual cell effectively. We presented two new
resource allocation schemes, the first converts the NP-hard mixed-integer resource allocation
problem into a continuous problem and
then finds an approximate solution, the second alternated
between the power allocation and channel allocation problems
when the channel allocation was carried out in a user-centric
manner. We proposed the use of hierarchical clustering
of the base-stations to form the virtual cells,
since with this technique changing the number of virtual cells only causes local
changes and did not force a recalculation of all the virtual cells
in the network. Finally, we presented numerical results
for the sum rate of these different techniques, and showcased where our newly proposed methods outperform existing techniques.
\bibliographystyle{IEEEtran}
%\bibliography{IEEEabrv,bib_research_abv}


\end{document}